# Effect of 200 MeV $Ag^{15+}$ ion irradiation on structural and electrical transport properties of $Fe_3O_4$ thin films

Ram Prakash[1*], R. J. Choudhary[1], Shailja Tiwari[1], D. M. Phase[1] and Ravi Kumar[2]

[1]UGC-DAE Consortium for Scientific Research, University Campus Khandwa Road, Indore-452017 India

[2]Inter University Accelerator Center, Aruna Ashif Ali Marg, New Delhi-110067 India

**Abstract**

Thin films of $Fe_3O_4$ have been deposited on single crystal MgO (100) and Si (100) substrates using pulsed laser deposition. Films grown on MgO substrate are epitaxial with c-axis orientation whereas, films on Si substrate are highly <111> oriented. Film thicknesses are 150 nm. These films have been irradiated with 200 MeV Ag ions. We study the effect of the irradiation on structural and electrical transport properties of these films. The fluence value of irradiation has been varied in the range of 5 x $10^{10}$ ions/cm$^2$ to 1 x $10^{12}$ ions/cm$^2$. We compare the irradiation induced modifications on various physical properties between the c-axis oriented epitaxial film and non epitaxial but <111> oriented film. The pristine film on Si substrate shows Verwey transition ($T_V$) close to 125 K, which is higher than generally observed in single crystals (121 K). After the irradiation with the 5 x $10^{10}$ ions/cm$^2$ fluence value, $T_V$ shifts to 122 K, closer to the single crystal value. However, with the higher fluence (1 x $10^{12}$ ions/cm$^2$) irradiation, $T_V$ again shifts to 125 K.

**Introduction**

Magnetite[1] ($Fe_3O_4$) is an important half metallic ferromagnetic material having 100% spin polarization at Fermi level. It has prospects in spintronic devices, magnetic storage and source as spin-polarized current injection. Some of its interesting properties are its high Curie temperature (~850 K), low electrical resistivity at room temperature and charge ordering at 120 K (Verwey transition[2]). At 120 K structural transition from room temperature cubic phase to monoclinic phase[1-2] and across this transition charge ordering also take place in the lattice. Thin films of $Fe_3O_4$ are generally grown on MgO substrate because MgO has a lattice constant half of that of $Fe_3O_4$. However, such a film is known to contain anti phase boundaries (APBs), which affect the electrical and magnetic properties enormously. Thin films of magnetite have been prepared by a variety of deposition techniques such as molecular beam epitaxy (MBE), reactive magnetron sputtering, pulsed laser deposition etc on different substrates like MgO, $MgAl_2O_4$, $α-Al_2O_3$, $SrTiO_3$, Pt, Si, GaAs etc[3-13] for their transport and magnetic properties. From these studies there is a consensus that the electrical transport properties vis a vis the magnetic properties of magnetite thin films are function of the nature of defects (internal or external) and defect density present in the film. For superior technological relevancies, compatibility between silicon substrate and $Fe_3O_4$ has been achieved. Recently, we have studied the compatibility of different substrates with $Fe_3O_4$. We observed that $Fe_3O_4$ thin films could be grown (111) oriented independent of substrates, when $Fe_3O_4$ has huge lattice mismatch with substrates using pulsed laser deposition techniques[14]. In another study[9] we have shown that $Fe_3O_4$ grow (111) oriented in different silicon substrate orientations ((100), (110) and (111)). There are no reports of SHI irradiation study of $Fe_3O_4$ films on Si substrates while there are a few reports for $Fe_3O_4$ on MgO substrates. Ravi kumar et al[13] reported SHI irradiation induced modifications in structural, magnetic and electrical transport properties in the $Fe_3O_4$ thin films grown on MgO substrates. Therefore, it is interesting to study the irradiation effect on oriented $Fe_3O_4$ thin films grown on Si substrates and compare these effects with previously reported irradiation effects on epitaxial thin films on MgO substrates. Swift heavy ion (SHI)[13] irradiation is an important tool, which is recognized to produce controlled defects, structural disorder, and strains in the thin film and to modify the physical properties of materials. In this paper we report the swift heavy ion irradiation effect on structural properties of $Fe_3O_4$ thin films on Si (100) and MgO (100) substrates.

**Experimental:**

The $\alpha$-$Fe_2O_3$ target used for the deposition of $Fe_3O_4$ thin films was prepared by the standard ceramic technique and was sintered at 950°C in air at atmospheric pressure for 24 hrs. $Fe_3O_4$ films were deposited on Si (100) and MgO (100) substrates by pulsed laser deposition (PLD) using a KrF excimer laser (Lambda Physik, Germany Model COMPEX-201, $\lambda$ = 248 nm). During the deposition, laser energy density was 1.8 J/cm$^2$ at the target and the substrate temperature was maintained at 450 °C. We did not perform any special treatment to remove the native oxide on surface of Si substrate, apart from the routine chemical cleaning using acetone and isopropanol in an ultrasonic bath for five minutes each. The distance between target to substrate was 5 cm. Prior to film deposition, the chamber was pumped down to ~ 10$^{-6}$ Torr. After the deposition, substrate was cooled down to room temperature in the same pressure as used during the deposition.

Films thicknesses were ~150 nm as measured by stylus profilometer. We irradiated these films with 200 MeV Ag ions using a 15 UD Tandem Accelerator, at Inter University Accelerator Centre (IUAC) New Delhi. The fluence value of irradiation has been varied in the range of 5 x $10^{10}$ ions/cm$^2$ to 1 x $10^{12}$ ions/cm$^2$. The ion beam was focused to a spot of 1 mm diameter and scanned over a 10×10 mm$^2$ area using magnetic scanner to irradiate the sample uniformly. To study the crystal structure of Fe$_3$O$_4$ thin film, X-ray diffraction measurements were carried out on rotating anode ( Cu) x-ray generator (Rigaku) with Cu-k$\alpha$ radiation source. The low temperature transport property was studied using standard four probe resistivity measurements.

**Results and discussions**:

It is well known that swift heavy ion (SHI) irradiation can produce controlled defect (point/cluster and columnar), modify the strain and transform the phase in the materials. When a high energy heavy ion passes through a material, it loses its energy mainly via two process; elastic process (nuclear energy loss) and inelastic process (electronic energy loss). The nuclear energy loss is dominant in low energy range (few hundreds of keV) and electronic energy loss is dominated in high energy range (10 MeV and above). In present case we used 200 MeV Ag ions on Fe$_3$O$_4$ thin films deposited on MgO and Si substrates by pulsed laser deposition. The electronic energy loss (Se), nuclear energy loss (Sn) and range of ion (Rp) have been calculated by using SRIM simulation program[15] and are found to be 25.94 keV/nm, 69.63 eV/nm and 12.3 μm respectively. From these values it is clear that Sn is three orders of magnitude smaller than Se and ion range is much larger than the film thickness (150 nm). To create columnar defects in materials certain threshold value of Se is required which is 36 keV/nm for bulk $Fe_3O_4$. In this case the Se is less than the required threshold value. For study of structural modification produced by SHI, we have performed XRD measurement. In Fig. 1 and 2 we show the XRD patterns of the $Fe_3O_4$ thin films on Si(100) and MgO(100) substrates. In the case of Si substrate the film is grown with preferred orientation in (111) direction with cubic structure. The film peaks are marked as F in the Fig 1 (a) and substrate peaks as Si. The irradiated films on Si substrate exhibit similar XRD pattern as that of the pristine film. We observed (111), (222), (333) and (444) reflections of $Fe_3O_4$ in pristine and irradiated films. From Fig.1(a) it is clear that (111) peaks of film has higher intensity than other peaks. Therefore, We have shown expanded plot of (111) peaks of pristine and irradiated films in Fig.1(b) and we have used this peak for further calculation of lattice parameter and grain size. We calculated the grain size ($D$) of the film using the Debye–Scherrer formula[16] given by

$$D = 0.94 \times \lambda/(B \cos \theta)$$

Where $\lambda$ is the wavelength of the x-ray source and $B^2=\Gamma^2-b^2$ in which $\Gamma$ is the full width at half maximum (FWHM) of an individual peak at $2\theta$ (where $\theta$ is the Bragg angle) and b is instrumental broadening. The lattice strain ($T$) in the material also causes broadening of diffraction peak, which can be represented by the relationship

$$T \tan \theta = (\lambda /D \cos\theta) - B.$$

The calculated lattice parameter and strain value of $Fe_3O_4$ thin films from the pattern are shown in table 1. It is clear that that the lattice parameter of all the samples is very close to the single crystal lattice parameter value of 8.39 Å. And strain remains almost constant in pristine and irradiated samples. However, $Fe_3O_4$ thin films grown on MgO (100) substrates are highly (h00) oriented as revealed by XRD pattern (shown in Fig.2). The peak positions (2h 0 0) of films coincide with the (h00) peaks of the MgO substrate as MgO has a lattice constant half of that of $Fe_3O_4$ in pristine and irradiated sample. Therefore, the possible changes in the lattice parameters of $Fe_3O_4$ films due to irradiation could not be resolved from their XRD patterns.

After structural characterization we have performed resistivity measurements for pristine and irradiated films deposited on Si and MgO substrates, shown in Fig.3. The pristine film on Si shows Verwey transition ($T_V$) at 125 K. After irradiation of a fluence 5 x $10^{10}$ ions/cm$^2$ Tv decreases to 122 K and at the fluence 1 x $10^{12}$ ions/cm$^2$ $T_V$ increased to 125 K, very close to the pristine sample. These results indicate that there is no appreciable change in $T_V$, consistent with our XRD results wherein we do not observe much change in lattice parameter and strain value even after irradiation. It is known from earlier reports[17] that nature of strain strongly influence the transport property and $T_V$ of $Fe_3O_4$ thin films. The activation energy of the film is calculated using Arrhenius equation ($\rho_H = \rho_0 \exp(-E_g/kT)$) beyond Verwey transition temperature. From the plot (not shown here) between $\ln(\rho)$ and 1/T we have calculated the activation energy, listed in table1, which is close to the reported data for $Fe_3O_4$ bulk value (58 meV)[18]. For the pristine $Fe_3O_4$ films on MgO substrates, the observed $T_V$ value is 120 K while after SHI irradiation of fluence value 1 x $10^{11}$ ions/cm$^2$, $T_V$ increases to 127 K. This increase in $T_V$ may be due to release of substrate induced strain, incorporated in the film during the deposition. After further increasing irradiation fluence value at 1 x $10^{12}$ ions/cm$^2$, $T_V$ does not increase further and it is again found at 127K for this fluence also. So after irradiation with a fluence of 1 x $10^{11}$ ions/cm$^2$, we may assume that the film was totally strain free.

**Conclusions**

In conclusion Fe$_3$O$_4$ thin films were deposited on Si (100) and MgO (100) substrates by PLD technique and irradiated by 200 MeV Ag ion in the range of 5 x $10^{10}$ ions/cm$^2$ to 1 x $10^{12}$ ions/cm$^2$. Before irradiation, X-ray diffraction study of pristine samples shows the spinel cubic structure of the films with preferential (111) orientation on Si (100) and (100) orientation on MgO (100) substrate. The resistivity measurements reveal that after irradiation T$_V$ does not changes much for films on Si substrates while films on MgO substrates show change in Tv from 120 K to 127 K.

**Acknowledgements**

Authers are thankful to Dr. P. Chaddah and prof. Ajay Gupta for encouragement. One of the authors (RP) thankful to CSIR, New Delhi, India for providing research fellowship.

**References:**

[1]. S. A. Wolf, D. D. Awschalom, R. A. Buhrman, J. M. Daughton, S. von Molnár, M. L. Roukes, A. Y. Chtchelkanova, and D. M. Treger, Science **294**, (2001) 1488.

[2]. E J W Verwey, Nature **144** (1939) 327.

[3]. D. T. Margulies, F. T. Parker, M. L. Rudee, F. E. Spada, J. N. Chapman, P. R. Aitchison, and A. E. Berkowitz, Phys. Rev. Lett. **79**, 5162 (1997).

[4]. S. K. Arora, R. G. S. Sofin, and I. V. Shvets *Phys. Rev.* **B 72**, 134404 (2005).

[5]. W. Eerenstein, T. T. M. Palstra, S. S. Saxena, and T. Hibma, Phys. Rev. Lett. **88**, 247204 (2002).

[6]. S. Kale, S. M. Bhagat, S. E. Lofland, T. Scabarozi, S. B. Ogale, A. Orozco, S. R. Shinde, and T. Venkatesan and B. Hannoyer, Phys Rev **B 64**, 205413 (2001).

[7]. D. M. Phase, Shailja Tiwari, Ram Prakash, Aditi Dubey, V. G. Sathe and R. J. Choudhary , J. Appl. Phys. **100**,123703 (2006).

[8]. S. B. Ogale, K. Ghosh, R. P. Shrama, R. L. Greene, R. Ramesh, and T. Venkatesan. Phys. Rev. **B 57**, 7823(1998).

[9]. S. Tiwari, R. J. Choudhary, Ram Prakash, and D. M. Phase, J. Phys.  Cond. Mat. **19**, 176002 (2007).

[10]. Y.X. Lu, J.S. Claydon and Y.B. Xu, Phys. Rev. B **70**,233304 (2004).

[11]. S. M.  Watts, C. Boothnan, S. van Dijken and J.M.D. Coey, J. Appl. Phys. **95**, 7465 (2004).

[12]. R.  J. Kennedy and P.A. Stampe, J.  Phys. D **32**,16 (1999).

[13]. Ravi Kumar, M. W. Khan, J.P. Srivastava, S. K. Arora, R. G. C. Sofin, R. J. Choudhary, and I. V. Shvets, J.Appl.Phys.**100** (2006) 033703.

[14]. Shailja Tiwari, Ram Prakash,  R. J. Choudhary and D. M. Phase, J. Phys. D: Appl. Phys.40 (2007) 4943.

[15]. J.F. Ziegler, J.P. Biersack, U. Littmark, The Stopping and Ranges of Ions in Solids, Pergamon, New York, (1985).

[16]. B. D. Cullity, "Elements of X-Ray Diffraction", Addison-Wesley, Reading, MA, (1972).

[17]. S.Jain , A. O. Adeyeye and C. B. Boothroyd, J. Appl. Phys. **97**, 10C312(2005).

[18]. A. Kazlowaski, R. J. Rasmussen, J. E. Sabol, P. Metacalf, and J. M Honig, Phys. Rev. B 48, 2057 (1993).

**Figure caption**

**Figure 1**: (a) X-ray diffraction pattern and (b) expanded plot of (111) peak of pristine and irradiated $Fe_3O_4$ films deposited on Si (100) substrates.

**Figure 2**: X-ray diffraction pattern of pristine and irradiated $Fe_3O_4$ films deposited on MgO (100) substrates

**Figure 3**: Resistivity versus temperature plot of pristine and irradiated $Fe_3O_4$ films deposited on (a) Si (100) and (b) MgO substrates.

Table1: Particle size, lattice parameter, strain, Tv and activation energy for pristine and irradiated $Fe_3O_4$ thin films on Si substrates.

| Sample | Particle size (nm) | Lattice parameter (Å) | Strain | Tv (K) | Activation Energy (Eg) (meV) |
|---|---|---|---|---|---|
| PRISTINE | 63 | 8.368 | .0015 | 125 | 61.8 |
| 5E10-IRR | 64 | 8.359 | .0016 | 122 | 58.5 |
| 1E11-IRR | 68 | 8.359 | .0013 | 124 | 62.9 |
| 5E11-IRR | 62 | 8.359 | .0014 | 123 | 64.5 |
| 1E12-IRA | 57 | 8.365 | .0015 | 125 | 61.3 |

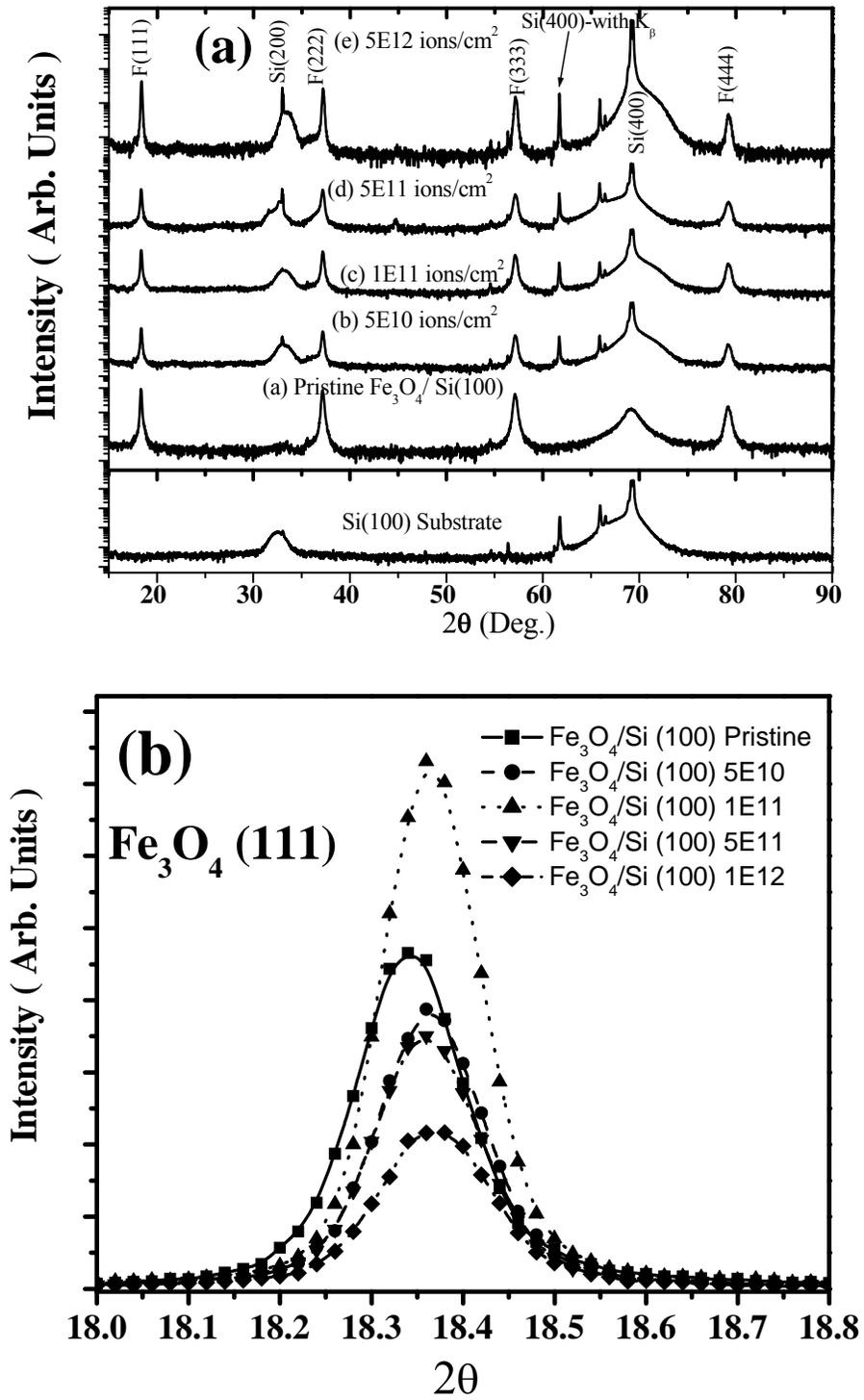

Figure 1

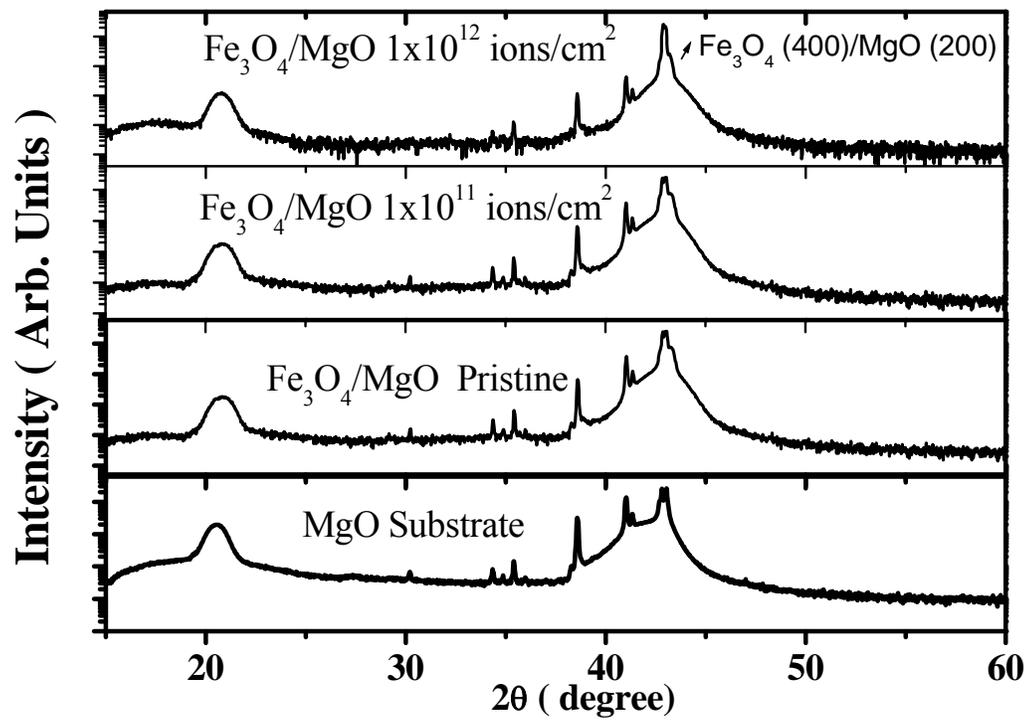

Figure 2

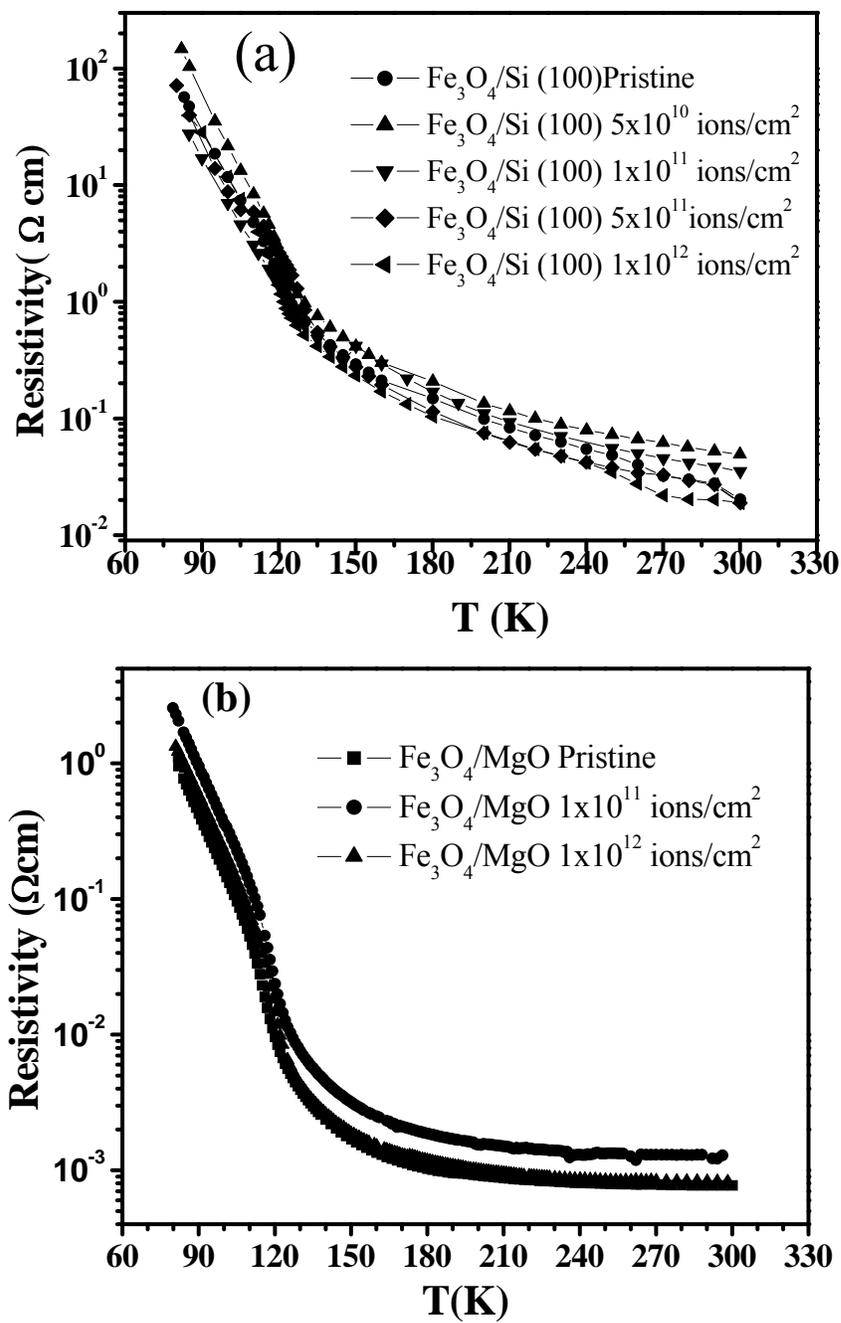

Figure 3